\documentclass{llncs}
\usepackage[utf8]{inputenc}
\usepackage{graphicx}
\usepackage{caption}
\begin{document} 

\title{Applying a Formal Method in Industry: a 25-Year Trajectory}
\titlerunning{Formal Method in Industry}  
%
\author{Thierry Lecomte\inst{1} \and 
David Deharbe \and Etienne Prun \and Erwan Mottin}
\authorrunning{Thierry Lecomte et al.} 
%
\tocauthor{Thierry Lecomte, David Deharbe, Etienne Prun, and Erwan Mottin}
\institute{ClearSy, 320 avenue Archimède, Aix en Provence, France\\
\email{thierry.lecomte@clearsy.com}}

\maketitle              

\begin{abstract}
Industrial applications involving formal methods are still exceptions to the general rule. Lack of understanding, employees without proper education, difficulty to integrate existing development cycles, no explicit requirement from the market, etc. are explanations often heard for not being more formal. 
Hence the feedback provided by industry to academics is not as constructive as it might be.

Summarizing a 25-year return of experience in the effective application of a formal method – namely B and Event-B – in diverse application domains (railways, smartcard, automotive), this article makes clear why and where formal methods have been applied, explains the added value obtained so far, and tries to anticipate the future of these two formalisms for safety critical systems.

\end{abstract}
\section{Introduction}
Formal methods and industry are not so often associated in the same sentence as the formers are not seen as an enabling technology but rather as difficult to apply and linked with increased costs.In \cite{DBLP:conf/fmics/Lecomte09}, the introduction of the B method and the Event-B language into several industrial development processes was witnessed with more or less success, even if new tools and new practices were available to ease acceptance in industry. At that time, these two formal methods had been backed by a number of research projects and non-trivial industrial applications. 

Almost 10 years later, after several real size experiments in diverse application domains, the situation has slightly evolved. Some standards, like the D0-178C for aeronautics, are now accepting formal methods in their certification process with sometimes some restrictions on the perimeter where they are applied (unit testing replaced by unit proof for example). The newborn ISO 26262 automotive functional safety standard is also recommending the use of formal methods during development. On the opposite side, the Common Criteria 3.1 standard (compared to its version 2.3) has decreased the need for formal methods that are now only required at level 6+ and higher (instead of 5+ previously) while the maximum security is reached at level 7 (EAL).
However, even if the standards have made some room for them, these methods haven't spread much out of the railway sphere as it might have been expected. Their usage though have slightly evolved over the years as a reaction to industry needs in direct relation with fierce international competition.

This article presents in a first chapter the different ways B and Event-B were used for modeling software, systems and data, and for proving static and dynamic properties. In a second chapter, new technology and techniques are presented. Their tight combination is expecting to converge to a new, more automated way of developing safety critical applications that are not restricted to the railways.

\section{Modeling}
\subsection{B for Software}
\label{sec:BforSoftware}
The B Method was introduced in the late 80’s to correctly design safe software. The main idea was to avoid introducing errors by proving the software while being built, instead of trying to find errors with testing after the software was produced.  

Promoted and supported by RATP, B and Atelier B\footnote{the tool implementing the B method} have been successfully applied to the industry of transportation, through metros automatic pilots installed worldwide. Paris Meteor line 14 driverless metro is the first reference application with over 110,000 lines of B models, translated into 86 000 lines of Ada. No bugs were detected after the proof was completed, neither at the functional validation, at the integration validation, and at the on-site testing, nor since the beginning of the metro line operation (October 1998).   

For years, Alstom Transportation Systems and Siemens Transportation Systems (representing a major part of the worldwide metro market) have been the two main actors in the development of B safety-critical software. Both companies have a product based strategy and reuse as much as possible existing B models for future metros. 
As an example, the Alstom Urbalis 400 CBTC (Radio communication based train control) equips more than 100 metros in the world, representing 1250 km of lines and 25 \% of the CBTC market. 

\subsubsection{Structure and metrics}
For such applications, B modeling is used for safety critical functions for both track-side (zone controller, interlocking) and on-board (automatic train pilot or ATP) software. The interlocking part has to avoid having two trains on the same track section. It computes boolean equations that represent the tracks status as seen from diverse sensors. The automatic pilot is mainly in charge of triggering the emergency brake in case of over-speed. It requires several functions such as the localization (\textit{where is the train ?}) that involve several graph-based algorithms, and the energy control which computes the braking curve of the train, based on the geometry of the tracks (in particular the positive and negative slopes). Data types used are: integer for the energy control, booleans for the interlocking and tables of integer for the tracks.

A typical ATP software model is made of one top-level function executed every cycle.

\begin{center}
\includegraphics[width=\textwidth]{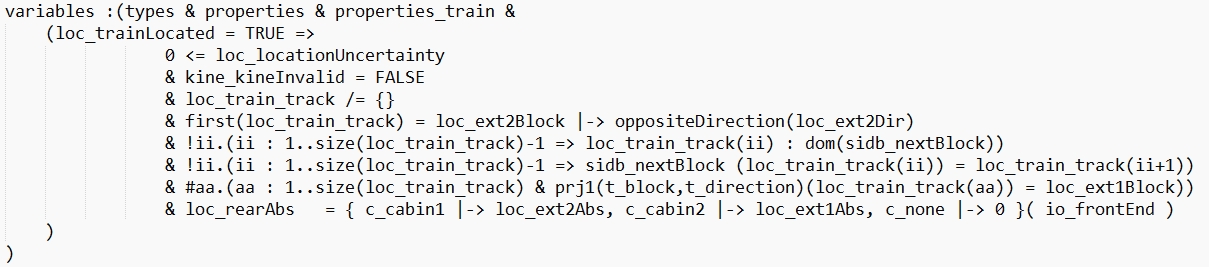}
\captionof{figure}{Example of a non-deterministic post-condition of a function}
\label{property-example}
\end{center}

The specification of this function (see figure \ref{property-example}) is non-deterministic and is expressed as a large "variables become such as" substitution. The specification of the function, contained in the post-condition, is sufficiently abstract and different from the implementation \footnote{ which contains the algorithm (statements, operation calls)} to avoid to prove the copy-paste from the specification to the implementation. This implementation imports 55 components. The complete B project is made of 233 machines (50 kloc\footnote{thousands lines of code}), 46 intermediate refinements (6 kloc) and 213 implementations (45 kloc), as well the handwritten code for non-safety critical parts (110 kloc). It also contains 3000 definitions reused among several components. 23,000 proof obligations are generated, 83\% of these of proved automatically, the remaining 17\% requires interactive proof. 3000 mathematical rules were added to ease the proof process, 85\% of these are proved automatically, the remaining 15\% requires human manual proof.  

To date, the biggest B software is a XML compiler enabling the execution of safety critical embedded applications by an interpreter. More than 300,000 lines of Ada code are generated from B models, for this SIL4 T3-compliant (EN50128) program \footnote{T3 means that the tool is able to generate a (faulty) binary program and as such requires a special attention in the safety process}. 300,000 lines do not represent the limit of the method as no bottleneck has been met until now. So the method is likely to scale up to larger, non-threaded software. A the other end of the scale, with platform screen doors controllers less demanding in term of computation, smaller applications are generated for both programmable logic controllers (PLC) and PIC32 micro-controllers, with a maximum of 64 KB in memory per software. 

\subsubsection{Organization and acceptance}
Since 1998, Atelier B has been slightly improved in order to obtain proven software more quickly:
\begin{itemize}
\item proof obligations (PO) contain traceability information (which parts of the B models have been used to obtain a PO), helping to better locate modeling errors and to improve modeling style
\item a model editor allowing to navigate models (abstraction, refinement) and operations (caller, callee)
\item a model editor merging model and proof (see figure \ref{editor-proof}) by displaying the number of proof obligations associated to any line of a B model, its current proof status (fully proved or not) and the body of the related proof obligations.
\item a framework to automatically prove and review user added mathematical proof rules, that generates a report for the safety case.
\end{itemize}

From a human point-of-view, usual organization requires a local guru acting as a technical referee (usually - but not necessarily - a PhD ) and a team of software engineers able to handle abstraction. Introductory B courses (B language, projects with B) and close support during the first months have been enough to set up development teams. The forthcoming MOOC on B\footnote{ https://moocs.imd.ufrn.br} and a dedicated YouTube channel for Atelier B practitioners would speed up the learning process.

\begin{center}
\includegraphics[width=\textwidth]{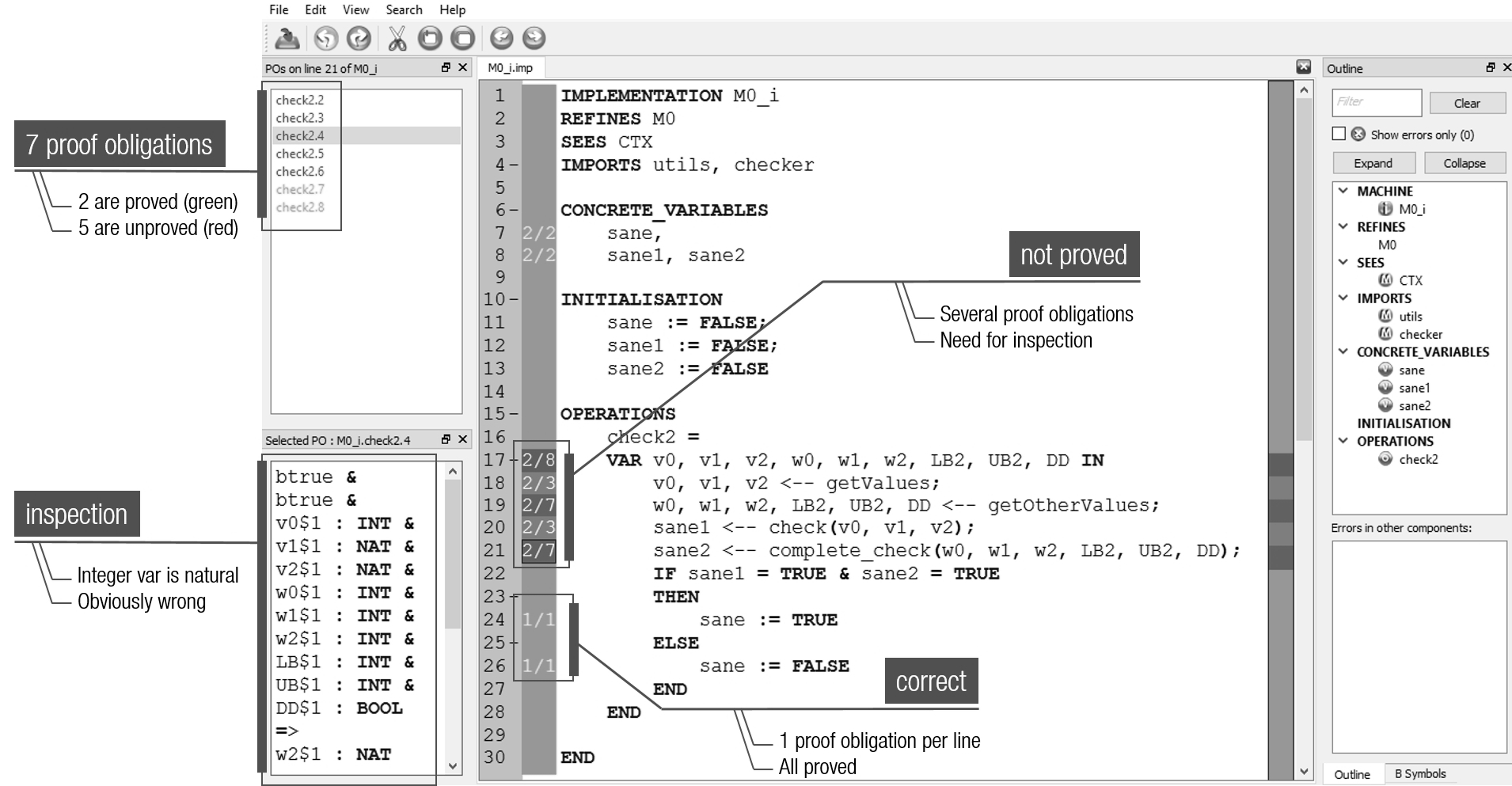}
\captionof{figure}{Text-based model editor combining proof information with modeling}
\label{editor-proof}
\end{center}

The B software development process is now well-oiled, accepted by certification bodies and several rail operators worldwide. Without being formally developed, Atelier B 3.6 was used for METEOR in 1998 while Atelier B 4.2/4.3 is used for Alstom Urbalis 400/500 product line. Atelier B 4.2 is at the core of the SIL4 certificate obtained for the platform screen-doors controller installed in 2017 in Stockholm (line Citybanan).

\subsection{B for Systems}
A broader use of B appeared in the mid ‘90s, called Event-B, to analyze, study and specify not only software, but also systems (system is here considered in its widest definition). It extends the usage of B to systems that might contain software, hardware and pieces of equipment, but also to intangible objects like process, procedure, business rule, etc.. In that respect, one of the outcome of Event-B is the proved definition of systems architecture and, more generally, the proved development of, so called, “system studies”, which are performed before the specification and design of the software. This enlargement allows one to perform failure studies right from the beginning, even in a large system development.

\subsubsection{Research and development}
\label{sec:Researchanddevelopment}
Several European projects were required to set-up Event-B, among them: 
\begin{itemize} 
\item MATISSE aimed at providing a first definition of the language, 
\item PUSSEE specifically aimed at hardware/software embedded systems, 
\item Rodin for the development of the eponymous platform and 
\item DEPLOY for its deployment in the industry. 
\end{itemize} 

Several system studies from diverse application domains (banking, air traffic control, defense, satellites, etc.) were initially performed with Atelier B  before naturally moving to the Rodin platform. The modeling of the Mazurkiewicz enumeration algorithm ands its proof during the project RIMEL \footnote{http://rimel.loria.fr/} was the perfect demonstration of the suitability of Event-B for small, distributed systems.
In 2008, during the certification for a smart-card microcircuit, Event-B was seamlessly integrated to Atelier B \footnote{because of the inability, at that time, for the Rodin platform to handle a model with 17 levels of refinement}. The supported language slightly differs from the one supported by Rodin but doesn’t restrict its usability regarding target applications. Several EAL5+ (CC2.3) and EAL6+ (CC3.1) certifications were performed in France, Germany and Spain, and functional specification were proved to comply with security policies. 

A follow-up project, FORCOMENT 
\cite{DBLP:journals/entcs/Benveniste11}, was initiated with STMicroelectronics and aimed at providing a proven path from specification to VHDL. Specific proof obligations were added to ensure a deterministic behavior. Resulting VHDL was quite different from the one developed manually (similar numbers of gates, but architecture more easily analyzable) and went successfully through product test benches. However the technology failed to find its audience because of:
\begin{itemize}
\item (the complexity of) the input formalism,
\item the necessity to specify the target system several tens of times (refinements) with different levels of detail,
\item the time and the number of iterations\footnote{Our maximum is 190 iterations and 5 major refactoring, many modifications having a slight impact on the structure of the model} to converge to a final model, 
\item the obligation to allocate our best practitioners to complete the duty.
\end{itemize}

\subsubsection{Flat specification}
Event-B was also used as a descriptive language for behavioral specification (flat specification, no refinement), mainly for document generation, structural analysis (dependencies among variables) and model animation with application in the automotive (enhanced diagnosis – Peugeot), in the defense (military vehicles integration testing scheduling – CNIM) and in the railways (platform screen doors preliminary studies – RATP). 

The main reason for not modeling with refinement was the complexity of the target systems and the level of detail required to perform an analysis that would have led to both practical and economical impossibilities (models too large to be handled by human modelers; too much effort to complete, if reachable). The Event-B models were sided by a dictionary containing natural language descriptions of the variables, events and substitutions, allowing for the automatic generation of document. Events were allocated to "sub-systems", allowing to analyze data-flows (see figure \ref{composys-graph}) between these sub-systems (where the variables are read/modified). 

A dedicated tool, Composys \cite{DBLP:conf/fm/Lecomte08}, was developed and maintained to support this approach until 2012. 

\begin{center}
\includegraphics[width=\textwidth]{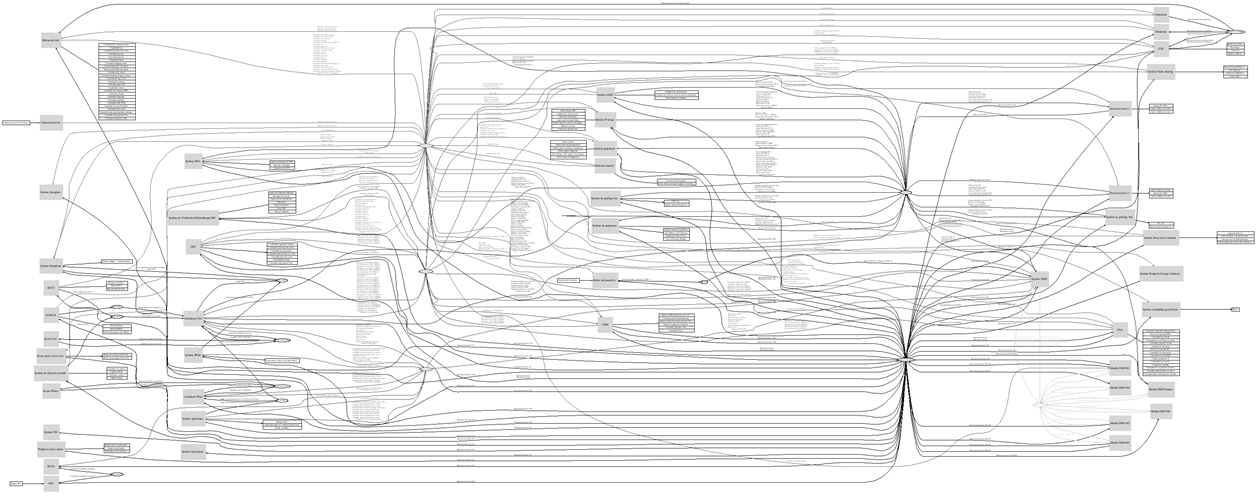}
\captionof{figure}{All the dependencies between the sub-systems of a military vehicle analyzed with Composys, and used for defining a non-trivial efficient integration testing policy. This drawing is for illustrating the complexity of the model.}
\label{composys-graph}
\end{center}

This approach was more aimed at finding ambiguities in the existing technical documentation, and at animating the specification than at proving a correct behavior and was finally abandoned.  

\subsubsection{Collection of separate models}
\label{sec:Collectionofseparatemodels}
Instead of developing a model of the whole signaling system 
, verbose, complex and not containing enough details \footnote{this demonstration requires for example to know the algorithm used for the odometer, to rediscover how the distance between signals and switches is computed based on the minimum curve radius, tunnel width, maximum slope, minimum train braking capability, etc.} to ensure a definitive conclusion on the safety of the system, another approach was tried.  The fundamental goal was to extract the rigorous reasoning establishing that the considered system ensures its requested properties, and to assert that this reasoning is correct and fully expressed. At system level, this rigorous reasoning involves the properties of different kind of subsystems (from computer subsystems to operational procedures), that the formal proof shall all encompass. Event-B is used to formalize the reasoning with a collection of separate models: each model is readable and understandable by a non-expert and doesn’t require to dig into hundreds of events and tens of refinement levels. This approach was used for the system formal verification for the CBTC of New York subway line 7 in 2012 and Flushing in 2014 (effort divided by two due to models reuse). It is now deployed in Paris for all the new automatic metro lines \cite{DBLP:conf/rssrail/Sabatier16}. Even if based on refinement, the formal modeling effort is now manageable (each model is one or two pages long) and only requires engineers able to reason (not our best practitioners any more).  The Event-B language as implemented in Atelier B in 2008 is still enough to support this modeling approach. 

\subsection{Formal Data Validation}
\label{sec:FormalDataValidation}
The verification of a behavior, based on Event-B system specification or B software specification, is achievable by semi-automated proof. However the verification of static properties of parameters (that tune the system or the software) against properties may turn out to be a nightmare in case of large data sets (10,000+ items) and complex relationships among data, as the built-in Atelier B prover is not able to handle them properly. In the early 2000's data validation in the railways \cite{DBLP:books/daglib/p/FalampinLLMP13} used to be entirely human, leading to painful, error-prone, long-term activities (usually more than six months to manually check constantly changing \footnote{CAD data is replaced by real plant data, topology is modified after in situ testing, etc.} 100,000 items of data against 1,000 rules). 

In 2003, this human process was made more formal while:
\begin{itemize}
\item formalizing data properties with the B mathematical language (set theory, first order logic)
\item generating a B machine containing the properties (the data model) and instantiated with the data to verify,
\item checking the correctness of the B machine
\end{itemize}

\subsubsection{Rules}
Properties, issued from international standards, national regulations, local practices, rail operator requirements, metro manufacturer constraints, are modeled as rules (see figure \ref{vdd-rule}). The clause \textit{WHERE} allows the selection and filtering of data \footnote{that could be stored in files like JSON, Xml, Ecxel, CSV, TXT, etc. }. The clause \textit{VERIFY} specifies the conditions expected for all filtered signals. In case the predicates of this clause are not verified, an error message is displayed for each signal found. 

\begin{center}
\includegraphics[width=\textwidth]{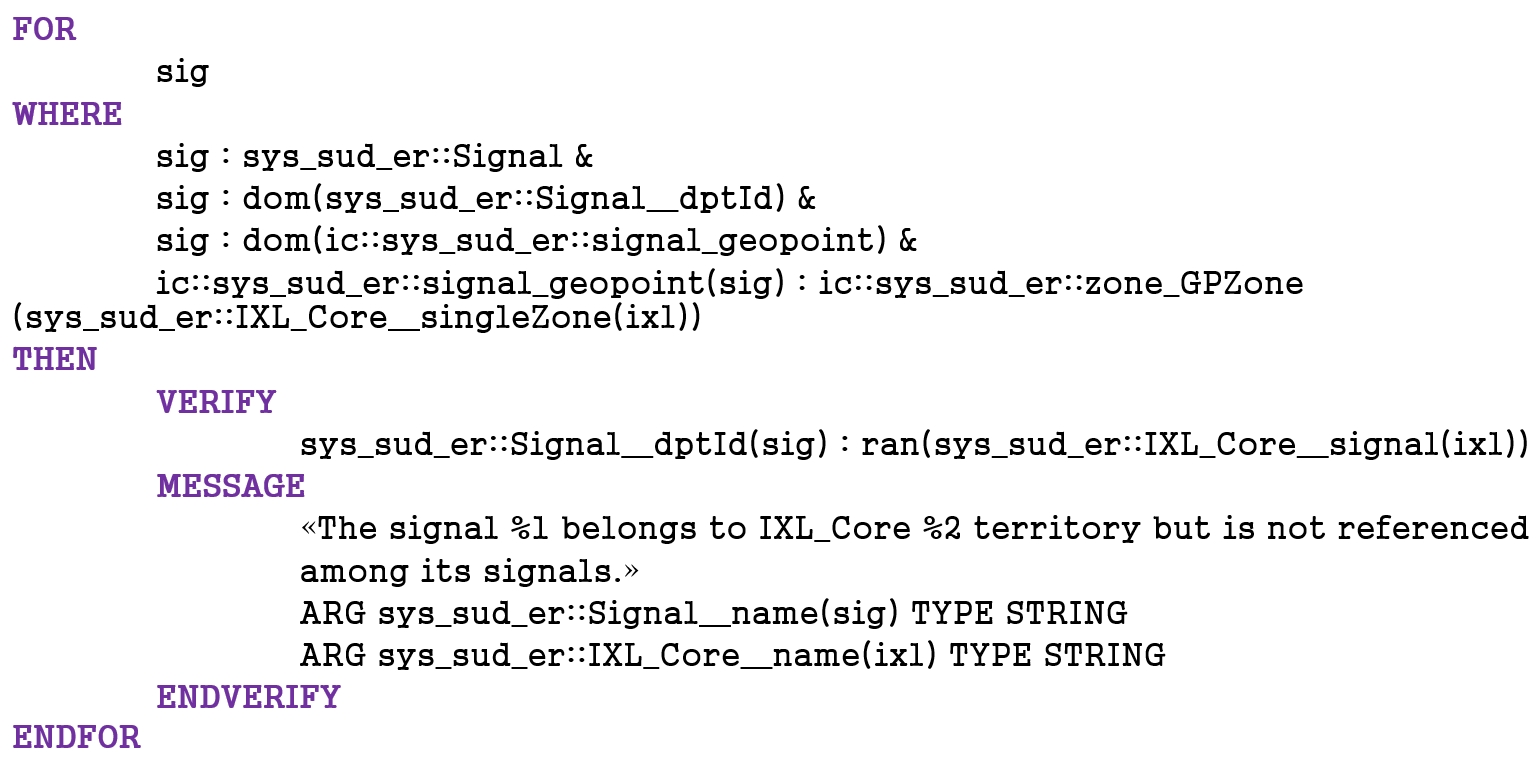}
\captionof{figure}{Example of verification rule. Signals belonging to an interlocking territory are searched; such signals have to be linked to this interlocking. If not, an error message is displayed for each faulty signal found.}
\label{vdd-rule}
\end{center}

Most of the rules fit in one page, but some rules are really large, up to 10 pages, as they embed several small steps or they contain a lot of implicit information. To ensure compliance with safety standard, rules have to be cross-read and tested by independent engineers. A specific testing environment has been developed to ease to set up of testing scenarios demonstrating that a rule triggers a KO conclusion for all error classes. 

\subsubsection{Deployment}
The PredicateB predicate evaluator was first used for checking the correctness. The PredicateB tool is a symbolic calculator able to manipulate B mathematical language predicates in order to animate a B formal model: constants and variables initial values are calculated, then operations are executed depending on enabling conditions and their substitutions. Symbolic values are scalars, sets, functions, etc. PredicateB has limited capabilities for non-deterministic computations and was replaced by ProB \cite{DBLP:conf/asm/HansenSL16}.
The ProB model-checker embeds several well performing heuristics for reducing search space (symmetry detection for example), is able to better handle non-deterministic substitution and to provide a more complete set of counter examples. It has been modified in order to produce a file containing all counter examples detected and slightly improved to better support some B keywords.

The major outcome of this decision to introduce formalities and to automate the verification \cite{DBLP:journals/corr/abs-1210-6815} was a dramatic reduction of the validation duration from about six months of human verification to some minutes of computation (if we set aside the time to formalize verification rules).
Since then the resulting tools (certified as T2 and T3 compliant, EN50128 standard) have been experimented with success \footnote{metro line fully and positively analyzed, results validated by certification body and independent expert} on several metro lines worldwide for different metro manufacturers. In this context, more than 2,500 rules have been developed, cross-verified and applied. The French Railways (SNCF) is going to deploy these tools for the main lines to check new interlocking parameters for the 10 coming years, requiring the development of 2,500 more rules. 

From a human point-of-view, usual organization requires engineers able to manipulate mathematical predicates and to understand railways signaling. A technical referee provides feedback and support on how to model certain tricky aspects like non-deterministic choices ("find a bijection such as ..."), quantified predicates, etc.
The verification process is well accepted by certification bodies and by several rail operators worldwide, and is ready to be deployed in other industries with safety-critical constraints. 

\subsection{Adoption by Industry}
From our experience, industry is not particularly interested in using formal methods except if it is required by the standards (1) or by the customers (2), or if it allows to speed up a process by an order of magnitude (3). 

\begin{table}
\begin{tabular}{|l|c|c|c|l|l|}
    \hline
    \textbf{Tool} & \textbf{B} & \textbf{E} & \textbf{D} &  \textbf{Usage} & \textbf{Availability}\\
    \hline
    {\small Atelier B} & X & X & & {\small modeling environment}  & {\small Free} \\
    & & & &  {\small 100+ automatic metro lines} & {\tiny http://www.atelier.eu/en} \\
    \hline
    {\small ProB} & X & X & & {\small model-checker} & {\small Free} \\
    & & & & & {\tiny https://www3.hhu.de/stups/prob} \\
    \hline
    {\small BMotionWeb} & X & X & & {\small model animator} & {\small Free} \\
    & & & & & {\tiny http://wiki.event-b.org/index.php/BMotion\_Studio} \\
    \hline
    {\small PredicateB} & & X & & {\small model animator} & {\small Free} \\
    & & & & & {\tiny https://sourceforge.net/p/rodin-b-sharp} \\
    \hline
    {\small PredicateB++} & & X & & {\small model animator} & {\small Proprietary (ClearSy)} \\
    \hline
    {\small Rodin} &  &  X & & {\small modeling environment} & {\small Free} \\
    & & & & & {\tiny http://www.event-b.org/} \\
    \hline
    {\small DTVT} &  &  & X & {\small data validation environment} & {\small Proprietary (Alstom)} \\
    & & & & {\small 20+ metro and tramway lines} &  \\
    \hline
    {\small Dave} &  &  & X & {\small data validation environment} & {\small Proprietary (General Electrics)} \\
    & & & & {\small Singapour metro line} & \\
    \hline
    {\small Ovado} &  &  & X & {\small data validation environment} & {\small Proprietary (RATP)} \\
    & & & & {\small Paris metro lines} & \\
    \hline
 \end{tabular}
 \caption{Summary of the main tools used during the last 25 years for industrial projects. B/E/D columns refer to B language (B), Event-B language (E) and formal data validation (D) supports.}
 \label{tab:tools}
 \end{table}
 
In our history, (1) is related to smartcard industry (§\ref{sec:Researchanddevelopment}), (2) is associated with Meteor/RATP (§\ref{sec:BforSoftware}) and with L7/NYCT (§\ref{sec:Collectionofseparatemodels}), while (3) is represented by the formal data validation (§\ref{sec:FormalDataValidation}).

In any case, a formal method without a proper tool support is useless. We have used several tools over the years (Table \ref{tab:tools}) that were applied in industrial settings. As such, formal data validation is much appreciated because as a V\&V tool, it doesn't impact the development cycle (on the contrary of B for software development) and the verification phase is a "push-button" activity (once the formal data model is completed).

\section{Convergence}
We have seen from the previous chapter that B and Event-B have matured over the last decade and are addressing well safety-critical industry topics \footnote{even if re-targeted to address more specific issues}, at system level, at software level, and at configuration level. However using a formal method is not enough to demonstrate safety. For example, a software can't be SIL4-compliant by itself, even if it is developed with B. The hardware executing it has to be considered, especially its failure modes, and a sound specification at system level has to be elaborated accordingly. A safety demonstration requires a lot of experience, skills, time and energy to complete successfully.

We present in this chapter several new features, linked with B, that are directly contributing to the safety demonstration and that would ease the development and the certification processes of safety-critical systems.
\subsection{Low Cost High Integrity Platform}
LCHIP \footnote{A short form of Low Cost High Integrity Platform} is a new technology, combining a complete software development environment based on the B language and a secured execution hardware platform, to ease the development of safety critical applications. It relies on several building blocks already used in certified railways products. 

LCHIP relies on a software factory that automatically transforms  function into binary code that runs on redundant hardware. The starting point is a text-based, B formal model that specifies the function to implement. This model may contain static and dynamic properties that define the functional boundaries of the target software.  

This formal specification is then refined automatically into a B implementable model. Transformation rules are applied to the specification to gradually replace abstract variables and substitutions with concrete ones.

The implementable model is then translated using two different chains:

\begin{itemize}
\item Translation into C ANSI code, with the C4B Atelier B code generator (instance I1). This C code is then compiled into HEX\footnote{a file format that conveys binary information in ASCII text form. It is commonly used for programming micro-controllers} binary code with an off-the-shelf compiler.
\item Translation into MIPS Assembly then to HEX binary code, with a specific compiler developed for this purpose (instance I2). The translation in two steps allows to better debug the translation process as a MIPS assembly instruction corresponds to a HEX line.
\end{itemize}

\begin{center}
\includegraphics[width=\textwidth]{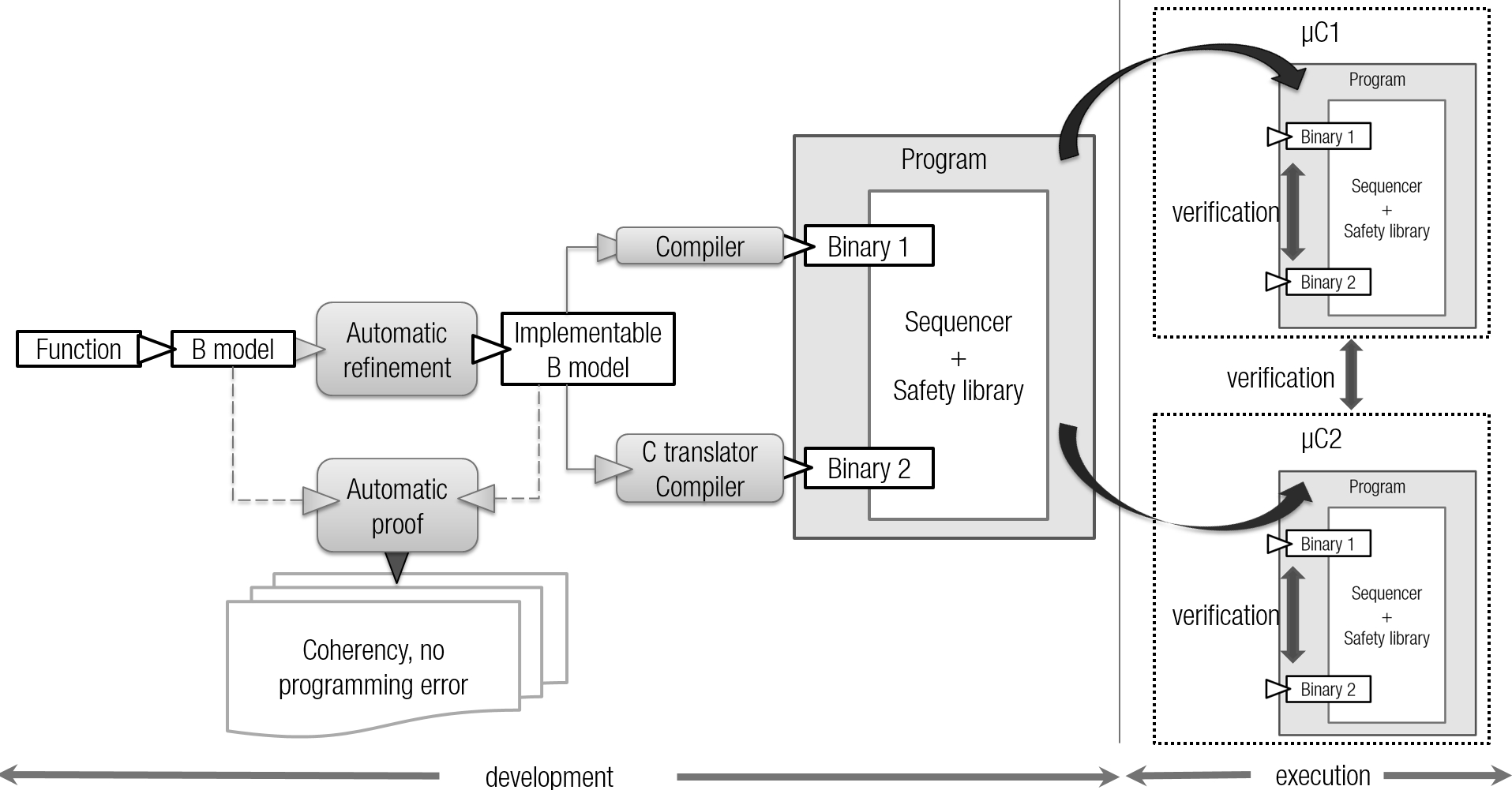}
\captionof{figure}{The safe generation and execution of a function on the double processor.}
\label{LCHIP-principe}
\end{center}

\subsubsection{Safety}
These two different instances I1 and I2 of the same function are then executed in sequence, one after the other, on two PIC32 micro-controllers. Each micro-controller hosts both I1 and I2, so at any time 4 instances of the function are being executed on the micro-controllers. The results obtained by I1 and I2 are first compared locally on each micro-controller then they are compared between micro-controllers by using messages. In case of a divergent behavior (at least one of the four instances exhibits a different behavior), the faulty micro-controller reboots. The sequencer and the safety functions are developed once for all in B by the IDE design team and come along as a library. This way, the safety functions are out of reach of the developers and can’t be altered.
The safety is based on several features such as the detection of a divergent behavior,
the detection of the inability for a processor to execute an instruction properly\footnote{all instructions are tested regularly against an oracle} and the ability to command outputs\footnote{outputs are read to check if commands are effective, a system not able to change the state of its outputs has to shutdown}. Memory areas (code, data for the two instances) are also checked (no overlap, no address outside memory range).
 
\subsubsection{Target software}
The execution platform is based on two PIC32 micro-controllers and provides an available power of 100 MIPS. This processing power is sufficient to update 50k interlocking Boolean equations per second, compatible with light-rail signaling requirements. The execution platform can be redesigned seamlessly for any kind of mono-core processor if a higher level of performance is required. 
Similar secured platforms are operating platform-screen doors in São Paulo L15 metro and in Stockholm City line. The Brazilian one has been recently certified at level SIL3 by CERTIFER on the inopportune opening failure of the doors.

The IDE provides a restricted modeling framework for software where:
\begin{itemize}
\item No operating system is used
\item Software behavior is cyclic (no parallelism)
\item No interruption modifies the software state variables
\item Supported types are Boolean and integer types (and arrays of)
\item Only bounded-complexity algorithms are supported (the price to pay to keep the refinement and proof process automatic)
\end{itemize}

The whole process, starting from the B model and finishing with the software running on the hardware platform, is expected to be fully automatic with the integration of the results obtained from some R\&D projects\footnote{to implement automatic refinement (ANR-RIMEL) and improve automatic proof performances (ANR-BWARE)}. In addition several in-house projects have helped to optimize the automatic refinement process by improving the refinement engine and by defining a subset of the B language, Simple B.

\subsubsection{Research and development}
LCHIP \cite{lecomte2016double} is developed by the eponymous French R\&D project.
It is aimed at allowing any engineer to develop a function by using its usual Domain Specific Language and to obtain this function running safely on a hardware platform. With the automatic development process, the B formal method will remain "behind the curtain" in order to avoid expert transactions.

As the safety demonstration doesn’t require any specific feature for the input B model, it could be handwritten or the by-product of a translation process. So several DSL are planned to be supported at once (relays schematic, grafcet) based on an Open API  (Bxml).
The translation from relays schematic is being studied for the French Railways with a strong focus on the feedback between DSL and B: in case of unproven B proof obligations, it is mandatory to exhibit its source in the DSL model.

The project reuses a number of building blocks such as the C4B\footnote{Atelier B C code generator} C code generator extended to support PIC memory model, and the B to Hex binary file in-house compiler supporting PIC32. 

The IDE will be based on Atelier B 5.0, providing a simplified process-oriented GUI. A first starter kit, containing the IDE and the execution platform, will be publicly released by the end of 2017. 

\subsection{Proof Support Advances}
\subsubsection{Proof Support in Atelier B}
A formal development demands that different aspects are verified using a mathematical proof. To this end, Atelier B produces automatically a number of proof obligations (POs). To assist the user in discharging POs, Atelier B has included a theorem prover since its inception. This "historical" theorem prover is an inference engine and an (extensible) rule database. It has been certified in the railway domain by expert review of both the inference engine and a core rule base. The architecture of the theorem prover is such that it can be used interactively, or automatically, at different force levels.

The user applies the theorem prover in batch to all the proof obligations, and is then left with a number of open POs. The remaining POs can be classified in three categories: valid, the theorem prover being unable to find the proof; unprovable, because the rule database is essentially incomplete; unprovable, because the user made a mistake in the formal development. 

The top priority of the user is to ensure that there is no mistake, i.e., 
there is no PO of the last category. Visually inspecting the POs is often enough
to detect most such errors, although there are also trickier mistakes that are
only uncovered in the course of an interactive proof. 

The user has then to discharge the unproved POs by interactive proof, and this is the
most time-consuming task in a formal development. The prover of Atelier-B supports a
number of commands to develop interactive proofs:  hypotheses selection, case split,
quantifier instantiation, equality rewriting, rule application, etc. A proof script is
successful when the proof obligation has been shown valid. One a script is successful,
it is saved in the project data base, and can be applied to other proof obligations.
Actually, a script is often successful for more than one PO. To improve scripting
capabilities and efficiency, the language has been enriched with pattern-matching
constructs that enable more general proofs. However, we feel that the interactive
proof process should be improved so that the user would only need to address
"interesting" goals and sub-goals that require some human insight. 

Since the specification language of the B method is undecidable, the user is allowed to 
write new rules to be taken into account by the inference engine. The risk of 
introducing inconsistencies is mitigated by two measures. The first measure consists in 
the inclusion of an alternative prover, based on tableaux, that is able to prove some 
of the rules automatically. The second measure applies to those rules that could not be 
proved automatically. It consists in the user writing a textual proof in natural 
language, that is then subject to validation by a third-party. 

In the past year, Atelier B support for PO verification has been improved with two 
different tools, addressing this issue at different levels:
\begin{description}
\item[iapa] (Interface to Automatic Proof Agents) for batch processing of POs;
\item[drudges] of the theorem prover for rapid processing of sub-goals in the interactive prover.
\end{description}
They are presented in turn in the following.



\subsubsection{iapa \label{sec:iapa}}
The iapa extension for Atelier B gives access to a number of third-party provers to discharge POs~\cite{DBLP:journals/corr/BurdyDP17}. In iapa, POs are not translated directly to the input format of these provers; instead the translation targets the format of a program verification platform that plays here the role of a gateway to such automatic provers, namely Why3~\cite{boogie11why3}. Each PO thus includes a prelude where the logic of the B expression language is formalized in Why3~\cite{mentre12abz}. The axiomatization of the B operators in Why3 has been fine tuned based on an industrial benchmark, resulting in significant improvement of the automatic proving capabilities in Atelier B on that benchmark~\cite{Conchon2014}.

\begin{figure}
\begin{center}
\includegraphics[width=\textwidth]{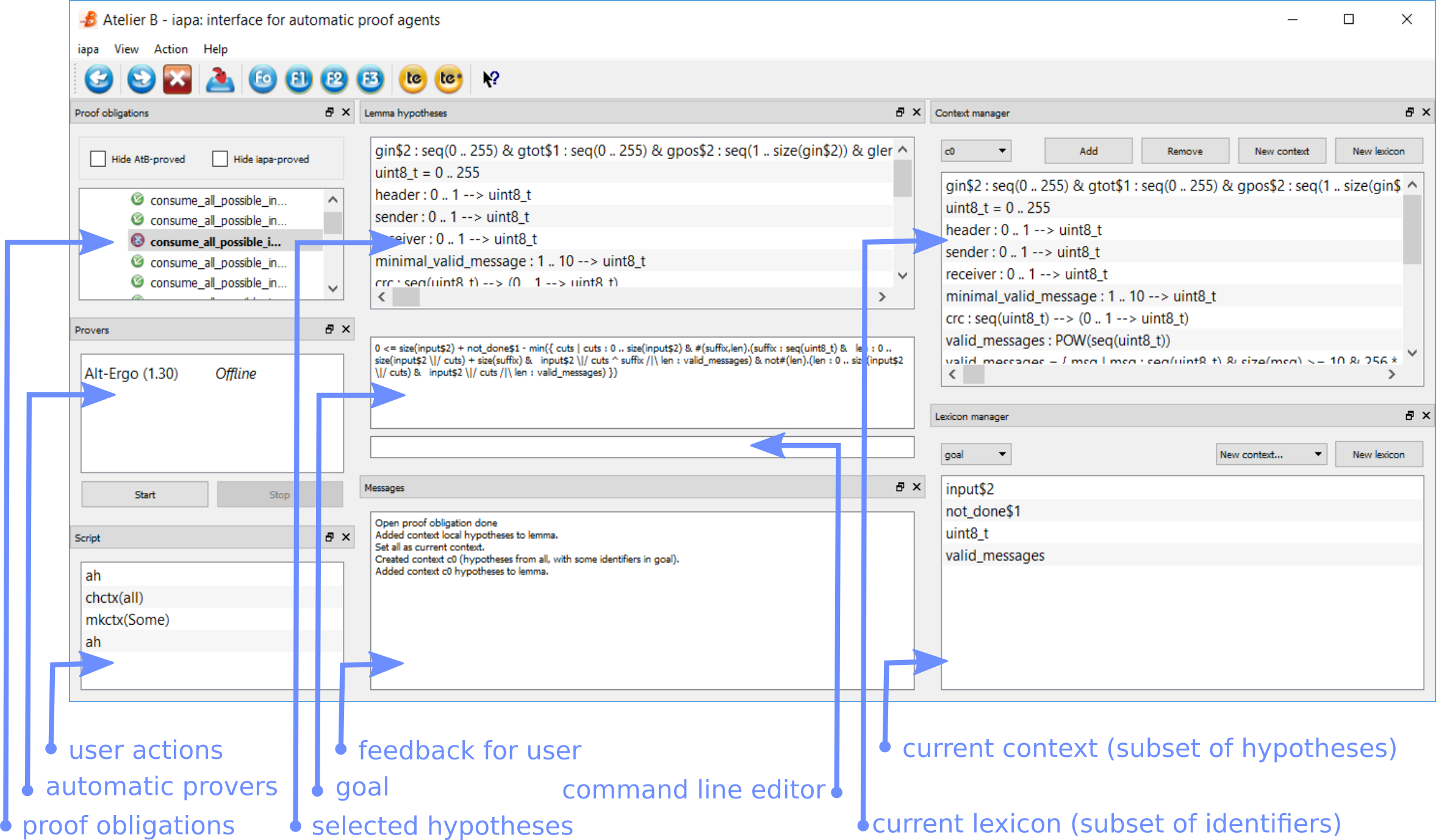}
\caption{An annotated screenshot of iapa}
\end{center}
\end{figure}

As the proof obligations are produced automatically, they include all the hypotheses that are in scope at the point the PO is concerned about. It is often the case that the validity of the goal only depends on a small number of such hypotheses. However, at times, provers are not able to identify these relevant hypotheses and end up lost in the proof search space. 

In order to address this issue, iapa includes a hypotheses selection functionality, where the user can identify a subset of the hypotheses, and only this subset is included in the proof obligation that is translated to Why3 and eventually processed by the provers. This functionality is available both through a graphical, point-and-click, interface and through a command line language. Subsets of hypotheses can be created according to the presence of some identifier or set of identifiers, then added to the proof obligation. Of course iapa also provides a function to extract a set of free identifiers from the goal or from some subset of hypotheses. These functionalities are built upon
two kinds of entities that the user can create and manipulate: contexts (subsets of hypotheses) and lexicons (subsets of identifiers).
Full details are available in~\cite{DBLP:journals/corr/BurdyDP17} ; iapa is part of Atelier B starting from version 4.5.

\subsubsection{Drudges of the interactive prover\label{sec:drudges}} The motivation for this functionality was born out of the feeling of frustration that the user of the interactive prover sometimes feels when she is faced with a seemingly trivial sub-goal, yet single command is able to discharge it. An example of such situation is when the current goal can be shown to be a consequence of the hypotheses using the theory of equality and propositional reasoning, but the terms involved are large or contain operators that get the automatic prover lost. 

\begin{figure}
\begin{center}
\includegraphics[width=.8\textwidth]{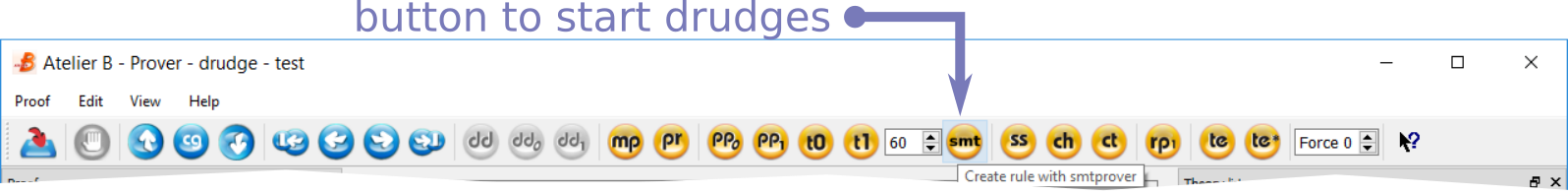}
\caption{Interface to the drudges in the window of the interactive prover \label{fig:drudges-before}}
\end{center}
\end{figure}

A general rule is that the less proficient the proof engine, the more efficient it is. So the rationale of the drudges of the interactive prover is to use automatic provers for simpler logics that are able to produce not only the result of the validity check, but also information on how they have reached their conclusion, and this information is then processed to produce guidance for the automatic prover of Atelier B. 

\begin{figure}
\begin{center}
\includegraphics[width=.8\textwidth]{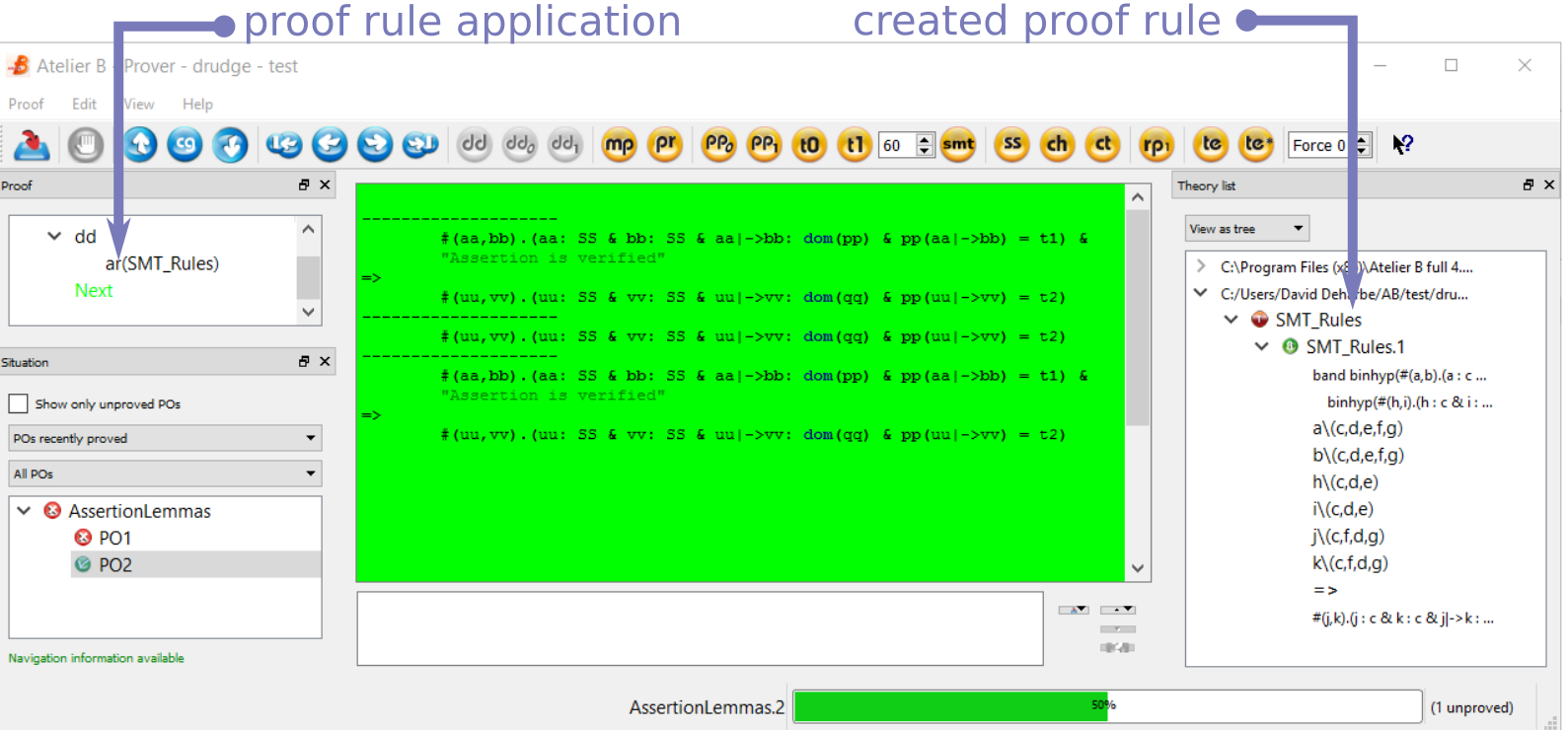}
\caption{State after the successful completion of the drudges \label{fig:drudges-after}: with a single click, a new rule has been created (right panel) and applied (left panel) automatically, discharging the goal.}
\end{center}
\end{figure}

Candidate drudges are provers that are either \emph{proof producing}, or at least able to generate a so-called \emph{unsat core}, i.e. a subset of the hypotheses that are actually used in the proof. Such functions have been standardized through at least two initiatives: TPTP~\cite{sutcliffe2009tptp} and SMT~\cite{BarFT-RR-15}. 
The drudges currently in the latter category only (veriT~\cite{bouton2009verit} and Z3~\cite{de2008z3}), as they implement the unsat core functionality. Given the unsat core, a proof rule for the Atelier B prover can be produced automatically, compiled and applied to the current goal. 
The drudges are available as a single click on a new button in the tool bar of the interactive prover (see figure~\ref{fig:drudges-before}). If the drudges are successful, the current goal is automatically discharged and the proof rule is added to the rule base of the component (see figure~\ref{fig:drudges-after}).

\section{Conclusion and Perspectives}
\subsection{Aimed at Industry}
Introducing formal methods in industry is difficult. We have experienced this situation with B in almost all industries, with a wide range of arguments:
\begin{itemize}
\item "we do not want to change of development cycle"
\item "we do not recruit PhD"
\item "formal methods work for train in 1-D, but planes flight in 3-D"
\item "trains and planes have professional drivers, but car drivers are mostly non-professional"
\item "we are not able to understand your deliverable" 
\item etc.
\end{itemize}

The real chance for the B-method was the very difficult development of the automatic speed control system for rapid transit railways in Paris, SACEM, in 1977, and the decision by the RATP to promote the B-method for the development of the first driver-less metro Meteor in 1993.

Several new usages at system-level and at configuration level have emerged over the last decade, scaling up to industry-strength deployments and offering new verification means with increased levels of confidence. These techniques allow to better manage complexity when dealing with large systems.
However, since 1994, B uses have been contained to a narrow scope of industrial software applications in the railways because of:
\begin{itemize}
\item the specific development cycle where unit and integration testing almost completely disappears,
\item the mandatory ability to handle abstraction for efficient modeling,
\item a specific code generator per target application to address hardware specifics.
\end{itemize}

The LCHIP technology, combined with improved proof performances and provers diversity, pave the way to an easier way of developing SIL4 functions (including both hardware and software). The platform safety being out of reach of the software developer, the automation of the redundant binary code generation process and the certificates already obtained for products embedding LCHIP building blocks, would enable the repetition of similar performances without requiring highly qualified engineers.
The hardware platform is generic enough to host a large number of complexity-bounded 
industry applications, with a special focus on the IoT and nuclear energy\footnote{In France several nuclear plants will have to be decommissioned in the coming years, requiring to develop supervision systems complying with current standards} domains.

\subsection{Challenges}

Safety-critical systems are certainly privileged targets when considering the application formal methods. The risk to injure or kill people may entitle to consider more easily "exotic" development, verification or validation means. With the raise of the IoT and the "connect-anything-to-anything" paradigm, security adds a new dimension to analyze and being able to model and prove at the same time safety and security properties could facilitate the acceptance of formal methods in the forthcoming standards releases.

Every industry has its own challenges. Based on our experience, our advice is to know and understand very well a particular application domain, especially its problems and imagine a usage of your formal method, even for a tiny / very specific scope\footnote{As such, LCHIP is a potential solution for small memory footprint safety-critical systems.}. Aim for the most automated process as industry is very fond of any "push-button" tool\footnote{Formal data validation is "usual" model-checking connected to a Domain Specific Language and traceability means to support certification.} .


\bibliographystyle{splncs03}
\bibliography{biblio}

\end{document}